\documentclass[]{aastex631}
\usepackage{amsmath}
\usepackage{pifont}

\newcommand{\eb}{\begin{equation}}
\newcommand{\ee}{\end{equation}}

\definecolor{rkka}{RGB}{219,66,32}
\definecolor{nsgreen}{rgb}{0.1,0.5,0.1}

\shorttitle{Orbital inclination of astrometric binaries}
\shortauthors{Makarov}

\begin{document}

\title{Orbital inclination of astrometric binaries and the dearth of face-on orbits in Gaia DR3 solutions}

\correspondingauthor{Valeri V. Makarov}
\email{valeri.makarov@gmail.com}

\author[0000-0003-2336-7887]{Valeri V. Makarov}
\affiliation{U.S. Naval Observatory, 3450 Massachusetts Ave NW, Washington, DC 20392-5420, USA}

\begin{abstract}
The orbital solutions for astrometric (unresolved) binary stars provided in the Gaia mission Data Release 3 reveal an obvious deficit of face-on orbits with line-of-sight inclinations close to 0 or $\pi$. This is shown to be an intrinsic mathematical feature of the orbit estimation technique involving the intermediate Thiele-Innes parameters, which are transformed to the Campbell geometric parameters. A direct condition equation defining the inclination angle via the Thiele-Innes values independently of the other orbital elements is used to investigate the origin of this near-degeneracy for face-on orbits. The emerging bias and correlation of inclination and semimajor axis are illustrated using Monte Carlo simulations for two specific template configurations representing  face-on and edge-on orbits. The results have significant impact on the interpretation and follow-up confirmation of Gaia-detected binary systems, including candidate exoplanets and brown dwarf companions.
 \end{abstract}

\section{Introduction} \label{int.sec}
Astrometric binary stars are detected from their apparent nonlinear motion in the local sky plane, which combines with the standard five-parameter astrometric model of stationary position, parallax, and proper motion of a chosen reference epoch. In the pre-space era of astrometry, a massive effort has been spent to characterize the orbital elements of the nearest and sufficiently well-resolved binaries using position measurements in the imaging mode or speckle interferometry \citep[][and references therein]{2001AJ....122.3472H, 2001AJ....122.3466M, 2024AJ....168..190T}. The observed trajectories in the sky plane projection are ellipses, whose centers coincide with the direction to the orbit's centers. The foci, on the other hand, are displaced from the true positions, so the distribution of mass cannot be determined from the projected trajectory. Traditionally, orbit characterization employed the methods of differential astrometry (i.e., position measurements of one companion relative to the other), while the basic five parameters (including the important parallax) were separately estimated via absolute astrometry or by indirect means.

The advent of space astrometry dramatically broadened the scope of astrometric exploration of binary stars by including unresolved pairs, where only one of the components is visible or the optical photocenter of a closely separated pair is observed. The orbit fitting becomes more complex involving all 12 unknowns (five astrometric parameters and seven orbital elements) within a single adjustment in the absolute astrometry regime \citep{1997ESASP1200.....E}. This nonlinear orbit adjustment is only possible when the orbital period is not much longer than the time span of the astrometric mission. Complete orbit quantification is not possible for long-period binaries, and the range of estimable physical parameters is limited to general statistical probabilities \citep{2024ApJ...972..186C}, rough statistical limits on the mass of invisible companions \citep{2005AJ....129.2420M}, and, in some cases, approximate mass ratios of nearby resolved systems \citep{2021AJ....162..260M}.

Orbital inclination $i$ has a special place in the set of orbital elements. It is needed to estimate the mass of the often invisible companion. The spectroscopic method using the radial velocity (RV) variations of the primary can only yield the minimum or projected mass of the companion $M_2\sin{i}$. The combination of the previously available precision RV series with the Gaia astrometric solutions is essential to pinpoint the secondary masses for numerous substellar candidates and separate real brown dwarfs from stellar systems with low-inclination orbits \citep{2023A&A...680A..16U}. Nearly face-on systems are of special significance because they are the main contaminants in RV-selected samples of candidate low-mass companions. The aim of this Letter is to reveal the underlying mathematical origin of the bias in inclination and semimajor axis estimation for astrometric binaries arising in the most commonly used technique of orbit quantification from only astrometric measurements in the tangential plane. A condition equation directly linking the cosine of inclination to the intermediate Thiele-Innes values is spelled out in Section \ref{ti.sec}, and its singularity for face-on orbits is discussed. In Section \ref{gaia.sec}, a brief review of the Gaia orbital solutions is given, which is known to show an anomalous distribution of estimated inclinations contradicting the expected isotropic distribution of orbital axes \citep{2023A&A...674A..34G}. Two specific orbits from Gaia are analyzed in Sections \ref{edge.sec} and \ref{face.sec}, with nominal parameters corresponding to nearly edge-on and nearly face-on orbits, respectively. Extensive Monte Carlo simulations reveal a catastrophic bias in the emerging estimates of inclination for the face-on configuration and a relatively moderate bias in the apparent orbit size. Discussion of the implications for the ongoing efforts to confirm exoplanets, brown dwarfs, and other interesting objects in the vast collection of Gaia astrometric binaries is proposed in Section \ref{end.sec}.

\section{Orbital inclination from Thiele-Innes parameters}
\label{ti.sec}
The equations for Thiele-Innes parameters are given in many textbooks and publications \citep[e.g., ][]{1978GAM....15.....H, 1985spas.book.....G}. They correspond to the components of the orbital position vector rotated from the frame defined by the orbital plane and the orbital momentum vector to the local celestial frame defined by the line of sight and the tangent plane. The specific functional form corresponds to the traditional for dynamical astronomy 3-1-3 sequence of Euler rotations by angles $-\Omega$, $-i$, and $-\omega$. The transformation is not completely unique, with known ambiguities emerging for zero eccentricity and face-on orbits. Are there alternative methods of geometrical orbit presentation that are free of internal degeneracies? The quaternion representation provides a healthy alternative minimizing the number of trigonometric functions and strongly nonlinear optimization loops \citep{2022MNRAS.513.2076M}, but this option has hardly been explored in the astronomical literature.

The derivation of the orbital elements $a$, $\Omega$, $\omega$, and $i$ in the Gaia DR3 solution is described by \citet[][Appendix A]{2023A&A...674A...9H}. It starts with the direct computation of $a$ using the following equation from \citet{Binnendijk+1960}:
\eb 
a=\sqrt{u/2+\sqrt{(u/2-v) (u/2+v)}},
\label{a.eq}
\ee 
where
\begin{align}
    u &= A^2+B^2+F^2+G^2 \nonumber\\
    v &= A\,G - B\,F
    \label{uv.eq}
\end{align}
The remaining parameters are computed in steps using rather complicated procedures, with $i$ emerging at the end after all the other parameters have been fixed. The formal uncertainties are estimated from the given formal errors of the along-scan position (abscissa) measurements via the rather involved approximate algorithm using the partial derivatives of the unknown values with respect to the abscissa \citep{2006ApJS..166..341G}. We note that this computation is based on the first-order approximation of the Taylor expansion and is adequate only for relatively small measurement dispersion and unbiased estimators. 

Here, we note, for the first time to my knowledge, that the inclination value can be directly computed from the TI elements via the following equation:
\eb 
\cos{i}+\sec{i}\equiv w =\frac{u}{v},
\label{cosi.eq}
\ee 
independently of other geometric angles and $a$. A related equation in a more complicated form is given by \citet{2022Icar..38315060E}. For a new variable $c\equiv \cos{i}$, the explicit solution of this equation furnishes two roots, $c_1=(w-\sqrt{w^2-4})/2$ and $c_2=(w+\sqrt{w^2-4})/2$. Both roots are real because $w$ can be positive and $w\ge 2$ or negative and $w\le -2$, which trivially follows from the pair of inequalities $(A-G)^2+(B+F)^2\ge 0$ and $(A+G)^2+(B-F)^2\ge 0$. Values of $w$ between $-2$ and $+2$ are not possible for any combination of TI values. However, only one of the roots yields a real-valued inclination angle $i$. Specifically, $c_1$ should be taken for a positive $w$ ($\cos{i}>0)$, and $c_2$ should be taken for a negative $w$ ($\cos{i}<0)$.

\begin{figure*}
    \includegraphics[width=0.48 \textwidth]{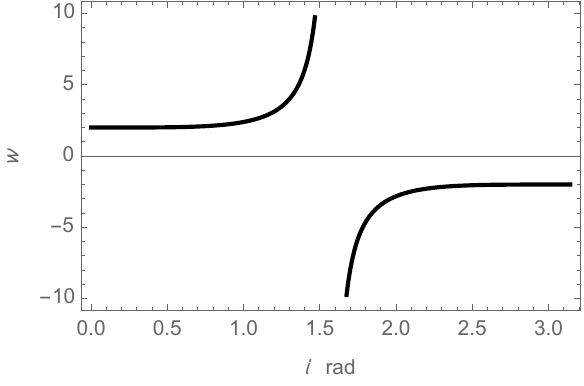}
    \caption{Dependence of $w=\cos{i}+\sec{i}$ on the inclination angle $i$. Note the flatness of the curves around $i\sim 0$ and $i\sim \pi$.}
    \label{w.fig}
\end{figure*}

Fig. \ref{w.fig} shows the functional dependence of $w$ on $i$. The curves become steep in the vicinity of $i=\pi/2$, which corresponds to an edge-on orbit, and quite flat toward the marginal values 0 and $\pi$ (counterclockwise and clockwise face-on orbits). Since the $w$ value is determined from observations with a certain dispersion of measurements, which is one-sided at 0 and $\pi$, this feature has a profound effect on the bias and confidence intervals of the estimated inclination values. Its origin is in the nearly singular dependence of $\cos{i}$ on $w$:
\eb 
\frac{d c}{d w}=\frac{c}{2\,c - w}.
\label{dc.eq}
\ee 
This derivative tends to infinity when $|c|\rightarrow 1$.

\begin{figure*}
    \includegraphics[width=0.47 \textwidth]{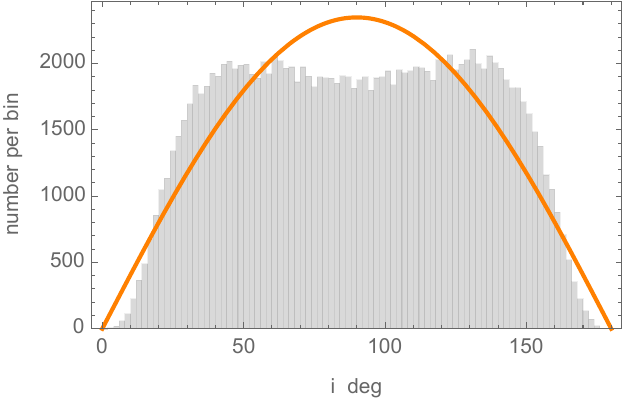}
    \caption{Histogram of estimated orbital inclination angles $i$ for 1.35 million Gaia DR3 astrometric binaries (gray-shaded area). The orange curve shows the expected PDF for random isotropic orientation of obits.}
    \label{i.fig}
\end{figure*}

\section{Orbital inclinations in Gaia DR3 solutions}
\label{gaia.sec}
The third Gaia mission release \citep{2016A&A...595A...1G, 2023A&A...674A...1G} included a dozen catalogs for various types of binary models under the common name non-single stars (NSS). In the Vizier database, this collection is available in the folder I/357. The largest catalog by far is I/357/tbooc, which includes 1.35 million objects. The entries have measured mean positions, proper motions, and parallaxes computed with the extended 12-parameter model. The Gaia solution provided the period, eccentricity, periastron epoch, and three or all four TI elements. The derived quantities include the projected semimajor radius $a_0$, inclination $i$, and the Euler angles $\Omega$ and $\omega$. This collection of candidate astrometric binaries surpasses the previous samples in size by orders of magnitude.

Examination of sample distributions reveals interesting facts about this collection. The histogram of the orbital periods $P$ is generally bell-shaped with large undulations of technical nature. If we account for the large deficit of periods around 1 yr and a smaller deficit around 2 yr caused by the poor conditioning of the 12-parameter model (essentially, a positive correlation between parallax and orbital excursion), the overall shape is well approximated by an Extreme Value distribution with location value 1.21 yr and scale parameter 0.6 yr. The histogram of the ``geometric" mass $m_{\rm geom}\equiv (a_0/\varpi)^3/P^2$, which is a substitute for the true total mass of the binary in $M_{\sun}$, shows a flat peak between 0 and 0.05 and a significant pileup or secondary peak at $\sim 0.004$ $M_{\sun}$. The latter may be seen as the presence of a large population of super-Jupiters orbiting Gaia stars with periods 1--3 yr. Objects with the largest $m_{\rm geom}$ are prime candidates for stellar mass black holes (StMBH), and the four candidates with $m_{\rm geom}>2.8$ $M_{\sun}$ include the three Gaia BH systems confirmed by follow-up observations. Objects with the smallest $m_{\rm geom}$ are in the focus of the exoplanet research community. While the overall empirical distribution of eccentricity is a bell-shaped function peaking at 0.25, the subset with low $m_{\rm geom}<0.001$ $M_{\sun}$ has a modal value around 0.4.

For this study, of special interest is the distribution of orbital inclination angles, which is shown in Fig. \ref{i.fig}. There is a striking difference between the empirical histogram (gray-shaded area) and the expected isotropic distribution of orbital axes (orange curve), which is proportional to $\sin{i}$. This anomalous histogram was discussed and analyzed in \citep{2023A&A...674A..34G}. The empirical PDF is concave with a shallow dip around $90\degr$, two symmetric bumps at $40\degr$ and $140\degr$, and a deficit of objects with inclinations around 0 and $180\degr$. The authors, using simplified Monte Carlo simulations with a fixed orbit size, eccentricity, and periastron argument set to zero, conclude that ``The apparent dearth of face-on configurations for Orbital solutions (Fig. 11) is therefore the consequence of extracting the Thiele-Innes parameters $A$, $B$, $F$, $G$ from noisy data". In this Letter, we will reveal the true origin of the feature, which is the singularity in the transformation of the Thiele-Innes parameters to the Campbell element $i$.

\section{An edge-on orbit}
\label{edge.sec}
We investigate a specific case when the nominal TI values correspond to a nearly edge-on orbit. As a template case for this option, we select a poorly investigated binary Gaia DR3 6422387644229686272 = WDJ201221.17$-$703642.44, which has been vetted as a white dwarf based on the results from Gaia DR2 \citep{2019MNRAS.482.4570G, 2019MNRAS.485.5573T}. The estimated (via stellar models for purely hydrogen and helium compositions) effective temperature is approximately 5800 K, and the mass with the updated photometry from Gaia DR3 is within $0.45\,M_{\sun}$ \citep{2021MNRAS.508.3877G, 2024A&A...682A...5V}. The object evaded the searches of nearby white dwarfs because of its relatively low temperature and reddish color, which probably indicate an old age.  With the orbital solution values of $19.5\pm 0.1$ mas for parallax, $a_0=2.62\pm 0.07$ mas for semimajor axis, and period $P=416\pm 2$ d, the mass of the companion comes up to $0.081\,M_{\sun}$. This suggests the companion to this white dwarf is a borderline brown dwarf. With the nominal $i=94.9\degr\pm1.6\degr$, the expected semiamplitude of radial velocity wobble from the companion is $3.69$ km s$^{-1}$. This should be detectable with modern-day spectrographs.

\begin{figure*}
    \includegraphics[width=0.47 \textwidth]{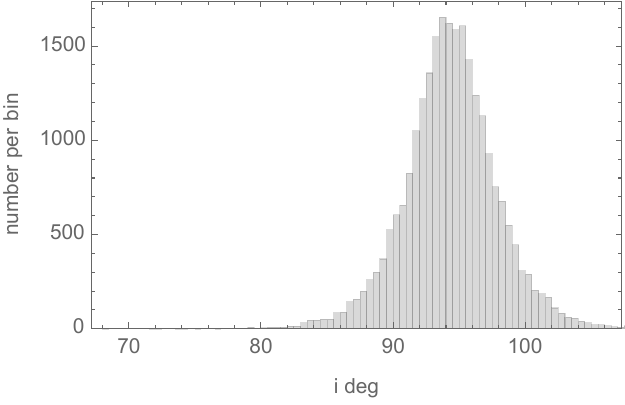}
    \includegraphics[width=0.47 \textwidth]{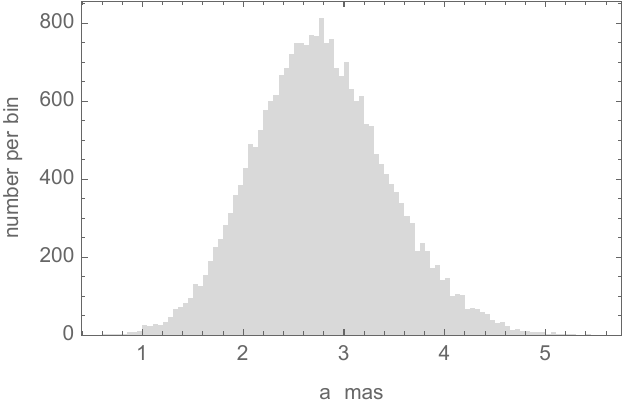}
    \caption{Histograms of inclination angles $i_{\rm mc}$ (left plot) and $a_{\rm mc}$ for the nominal set of TI values from the Gaia DR3 solution for Gaia DR3 6422387644229686272 = WDJ201221.17$-$703642.44, which corresponds to a nearly edge-on orbit ($i=94.9\degr$ and $a=2.62$ mas). Each MC sample includes 25,000 simulations.}
    \label{edge.fig}
\end{figure*}

For verification, Eqs. \ref{a.eq} and \ref{cosi.eq} are used to compute $a_0$ and $i$ directly from the nominal TI values specified in the Gaia catalog. The results are identical to the cataloged values: $i=94.904688\degr$ and $a=2.6215323$ mas. Instead of the imprecise and elaborate procedure to estimate the errors (or confidence intervals) of these values used by the Gaia consortium, we perform Monte Carlo simulations capitalizing on the fastness and accuracy of these equations. In all, 25,000 random realizations of the TI set $\{A,B,F,G\}$ are generated by normally distributed random numbers with the nominal values as the means and the specified formal errors as the standard deviations. For each realization, a pair of orbital elements $\{i,a\}$ is computed. The resulting 2D sample distribution can be used to estimate the maximum likelihood values of these elements and the robust estimates of their uncertainty. The histograms of Monte Carlo (MC) samples are shown in Fig. \ref{edge.fig}. The bell-shaped histograms, which are in fact not Gaussian because of the enhanced tails, are fairly symmetric around their mean values $i_{\rm mc}=94.38\degr$ and $a_{\rm mc}=2.77$ mas. The means are not too different from the nominal solution values in this case. A significant difference is found for the dispersion of the Monte Carlo sample in inclination, which comes up to $3.65\degr$ against the estimate $1.6\degr$ given in the Gaia solution. The MC estimate of the standard deviation of $a_0$ at $0.67$ mas is even more discrepant with the Gaia value $0.07$ mas. Thus, for a nominally edge-on configuration at a comfortably high signal-to-noise ratio, we find strongly underestimated uncertainties of orbit size and inclination angle, but no evidence of a significant bias.

\begin{figure*}
    \includegraphics[width=0.47 \textwidth]{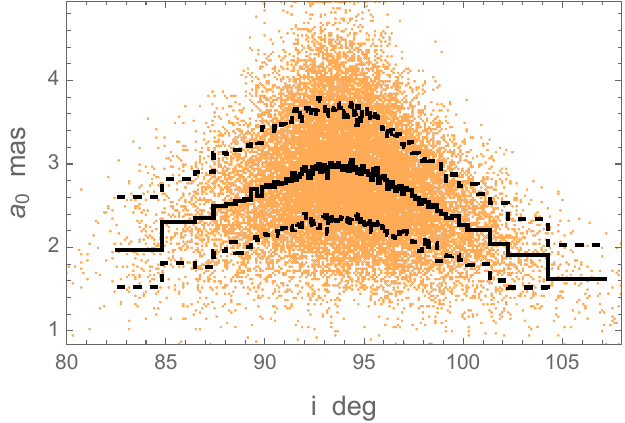}
    \caption{2D distribution of 25,000 MC-simulated solutions for apparent orbital size $a_0$ and inclination $i$ obtained from the nominal Thiele-Innes values and their formal uncertainties in Gaia DR3 (orange dots). The median values of these parameters in bins of 100 trials sorted by $i$ are shown with the solid stepwise black curve. The $\{0.16,0.84\}$ quantile values for the same bins are represented by the black dashed lines.}
    \label{ia1.fig}
\end{figure*}

The published orbital solutions do not provide any information about correlations between the orbital parameters. Reliable uncertainties and covariance of $a$ and $i$ are especially important for risk assessment and target selection in spectroscopic follow-up observation of candidate low-mass companions. The MC simulation of nominal high-inclination solutions starting with the given TI values and their uncertainties can partially address this problem. Fig. \ref{ia1.fig} shows the 2D distribution of 25,000 trial MC solutions as a cloud of orange points. Obviously, this distribution is not axially symmetric, and the normally assumed bivariate Gaussian density is a bad hypothesis even for this high-SNR case. The solid black curve represents the binned median $a_0$ versus median $i$, and the dashed lines show the $\{0.16,0.84\}$ quantile interval (the robust alternative to $\pm 1\sigma$ interval). The mean value of $i$ corresponds to the statistically highest estimates of $a_0$, while possible solutions away from $94\degr$ result in smaller orbit size estimates. Using the empirical MC results, a user can accurately compute the probability that the nominal configuration corresponds to a radial velocity signal above a certain threshold.
 
\section{A face-on orbit}
\label{face.sec}
The deficit of face-on orbits is so pronounced in the Gaia DR3 solution that we find only four stars (out of 1.35 million) in the general catalog with nominal $i<3\degr$. One of these stars is Gaia 1988288559178163840 with the cataloged $i=2.63\degr$, $a=0.467$ mas. Using the given values of TI elements, we generate 25,000 MC simulations perturbing each element by an independent random number from a Gaussian distribution with the specified standard deviation. Using Eqs. \ref{a.eq} and \ref{cosi.eq}, identical solutions are obtained for the ``exact" values of TI elements. The output of MC simulations paints a drastically different picture from the previous example.

\begin{figure*}
    \includegraphics[width=0.47 \textwidth]{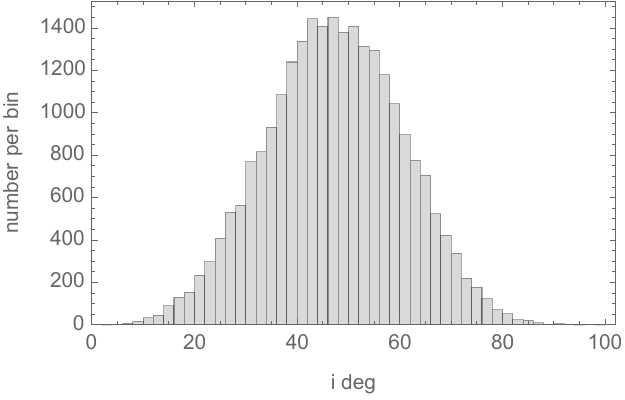}
    \includegraphics[width=0.47 \textwidth]{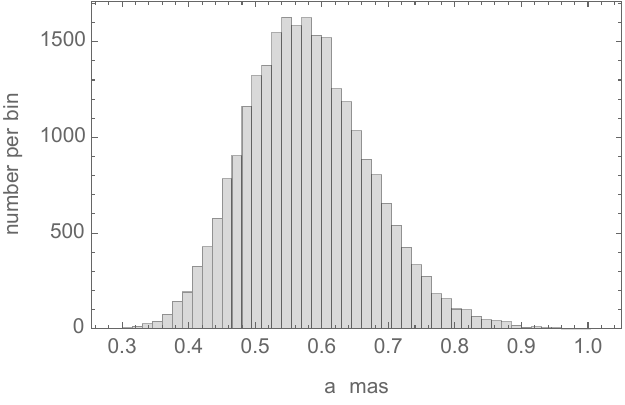}
    \caption{Histograms of inclination angles $i_{\rm mc}$ (left plot) and $a_{\rm mc}$ for the nominal set of TI values from the Gaia DR3 solution for star 1988288559178163840, which corresponds to a nearly face-on orbit ($i=2.63\degr$, $a=0.467$ mas). Each MC sample includes 25,000 simulations.}
    \label{face.fig}
\end{figure*}

Fig. \ref{face.fig} shows the MC-generated distributions of inclination and semimajor axis.
Although the sample distributions are still bell-shaped, the emerging values are heavily biased with respect to the nominal solution. The empirical mean(std) values from the MC sample are $47.3\degr(13.3\degr)$ for inclination and $0.579(0.094)$ mas for semimajor axis. The bias in $a$ is larger that $1\sigma$, while all the 25,000 MC estimates miss the nominal $i$ completely. This means that a binary with the true inclination and TI values as given, corresponding to a nearly face-on orientation, would almost certainly end up with a high estimate of inclination around $45\degr$, and a significantly larger $a$ than the correct value. A 2D plot of the MC $\{a,i\}$ estimates reveals that these parameters are also statistically correlated, so that a higher inclination is associated with a longer $a$. The Spearman's correlation is computed at 0.4. 

\section{Conclusions and Discussion}
\label{end.sec}
A more direct and simple method of computing the orbital inclination angle $i$ from a given set of four TI values is proposed via Eq. \ref{cosi.eq}. The mathematical problem in question involves four condition equations and four unknowns and is almost everywhere exactly determined (except for the well-known degeneracy between $\omega$ and $\Omega$ at $e=0$). However, the new method is quite useful for performing massive Monte Carlo or MCMC simulations of the nonlinear transformation of TI values to the corresponding orbital elements, which reveal the hidden biases, correlations between the solution results, and robustly estimated uncertainties.

Eq. \ref{cosi.eq} is sufficient to compute a unique value of $i$ between 0 and $\pi$, which is identical to the results obtained with other algorithms, including the one used by the Gaia processing team. Being independent of the other orbital parameters, it contains complete information about the distribution of estimated $i$. To reveal this information, a Monte Carlo simulation can be performed for each set of TI values. The character of the characteristic function $w=\cos{i}+\sec{i}$ (Fig. \ref{i.fig}) explains the shape of the sample distribution of Gaia-estimated inclinations with a pronounced dearth of values close to 0 and $\pi$, i.e., of face-on orbits. In this domain, $w$ is nearly constant for a wide range of $i$. This makes the gradient $d \cos{i}/d w$ practically singular, which, combined with the one-sided dispersion of $w$ due to measurement errors, results in a catastrophic bias. 

The numerical experiments presented in \citep{2023A&A...674A..34G} have shown that the magnitude of the bias in $\cos{i}$ depends on the SNR of the intermediary TI estimates. These results did not reveal the cause of the face-on deficit, nor the related bias in the estimated orbit size and the intrinsic correlation of $i$ and $a$. The problem facing the specialists and users of astrometric binary solutions is therefore of more immediate mathematical kind than to just collect more measurements and improve the SNR of the TI elements. We have to look for viable alternatives to the TI method of orbit quantification. The practical convenience of linear equations linking TI parameters with star abscissae should perhaps be sacrificed to avoid the singularity of the inclination solutions.

The conclusions of this study are relevant for the ongoing effort to verify and confirm the most interesting Gaia-detected binaries. The two analyzed cases represent the range of geometric configurations with nominally near-face-on and edge-on orbits. Gaia DR3 6422387644229686272 = WDJ201221.17$-$703642.44 is a recently detected field white dwarf at the bottom of the cooling sequence with possibly a brown dwarf companion. It is used in this study as a favorable template case of nearly edge-on solution with high SNR. MC simulations were used to probe for the actual dispersion and bias at the TI to $\{i,a\}$ transformation step. These simulations reveal a relatively small bias in $i$ and $a$ but a critically underestimated uncertainty of $a_0$ listed in the catalog ($0.67$ mas against $0.07$ mas). The distribution of MC trials demonstrates a complex, asymmetric structure and a pronounced correlation between the estimated orbit size and inclination (Fig. \ref{ia1.fig}). It shows that the expected radial velocity amplitude for this object is more likely to be smaller than the estimated value from the nominal solution.

To investigate the empirical deficit of low inclination angles, similar MC simulations were performed for one of the few stars with a nominal face-on orbit. The available TI values were used as the true means, and the TI values were perturbed by a Gaussian random number according to the given standard deviations. The emerging sample distribution of $i$ is catastrophically biased, so the assumed true value is practically never realized. The simulated sample mean inclination is $47.3\degr$, which matches very well with one of the symmetric peaks in the histogram of estimated $i$ (Fig. \ref{i.fig}). Therefore, this enhancement of intermediate inclination angles includes biased face-on configurations. The bias is not an error in the Gaia processing of astrometric binaries but a mathematical feature intrinsic to the TI method. It is caused by the singularity of the explicit equation representing the transformation of the TI parameters to $i$. Unfortunately, it is not possible to back-engineer the maximum likelihood $\{i,a\}$ parameters given a set of TI values and estimated bias. The users of the astrometric binary catalog should keep in mind the possibility of a large error in the estimated $i$ and a positively correlated, but relatively smaller, bias in $a$. 

The intrinsic poor condition of inclination reconstructed from TI values for face-on configurations is a complicating factor for the efforts to verify the trove of new astrometric binaries detected by Gaia. \citet{2023AJ....165..266M} investigated a sample of six candidate exoplanet systems from the vast Gaia DR3 collection. While three systems turned out to be double-lined spectroscopic binaries (an anticipated source of false positives due to the photocenter effect), two other objects were confirmed, albeit with greatly smaller RV amplitudes than the Gaia prediction. In total, half of the systems with available RV data revealed inconsistencies between the magnitude of RV signals and the Gaia-estimated inclinations. These cases are manifestations of the bias in inclinations revealed by our analysis. The same feature fully explains the inverted S shapes in the comparison of inclinations derived from Campbell and TI parameters for astrometric-spectroscopic binaries \citep[][their Figs. 7 and 10]{2005A&A...442..365J}.

\section*{Acknowledgements}
This work has made use of data from the European Space Agency (ESA) mission
{\it Gaia} (\url{https://www.cosmos.esa.int/gaia}), processed by the {\it Gaia}
Data Processing and Analysis Consortium (DPAC,
\url{https://www.cosmos.esa.int/web/gaia/dpac/consortium}). Funding for DPAC
has been provided by national institutions, in particular the institutions
participating in the {\it Gaia} Multilateral Agreement. This research has made use of the VizieR catalogue access tool, CDS,
Strasbourg, France \citep{10.26093/cds/vizier}. The original description 
of the VizieR service was published in \citet{vizier2000}.

\bibliography{main}
\bibliographystyle{aasjournal}

\end{document}